\documentclass[12pt,a4paper]{article}
\usepackage[hscale=0.8,vscale=0.85]{geometry}
\usepackage{graphics}
\usepackage{stmaryrd}
\usepackage{amsfonts}
\usepackage{mathrsfs}
\begin{document}

\title{From Prigogine to Raychaudhuri}

\author{Minyong Guo$^1$\thanks{Email: minyongguo@mail.bnu.edu.cn }, Yu Tian$^{2,3}$\thanks{Email: ytian@ucas.ac.cn}, Xiaoning Wu$^4$\thanks{Email: wuxn@amss.ac.cn}, Hongbao Zhang$^{1,5}$\thanks{Email: hzhang@vub.ac.be}\\
\footnotesize $^1$Department of Physics, Beijing Normal University,
\\\footnotesize Beijing 100875, China
\\\footnotesize $^2$School of Physics, University of Chinese Academy of Sciences,
\\\footnotesize Beijing, 100049, China
 \\\footnotesize $^3$Shanghai Key Laboratory of High Temperature Superconductors,
 \\\footnotesize Shanghai 200444, China
 \\\footnotesize $^4$Institute of Mathematics, Academy of Mathematics and System Science, Chinese Academy of Sciences,
 \\\footnotesize Beijing 100190, China
 \\\footnotesize $^5$Theoretische Natuurkunde, Vrije Universiteit Brussel, and The International Solvay Institutes,
 \\\footnotesize Pleinlaan 2, B-1050 Brussels, Belgium}

\date{}

\maketitle

\begin{abstract}
It is highlighted by Prigogine that there are two additional universal behaviors associated with the entropy production rate besides the four laws of thermodynamics. One is that the entropy production rate decreases when the system approaches the steady state, and the other is that the entropy production rate reaches its minimal value at the steady state. Motivated by the black hole thermodynamics and AdS/CFT correspondence, we resort to Raychaudhuri equation to prove that these two universal behaviors are also obeyed by the black hole entropy. In particular, our result together with the four laws of black hole thermodynamics further indicates that the holographic gravity should be universal, not restricted only within the context of AdS/CFT correspondence.
\end{abstract}

\section{Introduction}
In 1973, Bardeen, Carter, and Hawking formulated the four laws of black hole mechanics, which bear a strong resemblance to the ordinary laws of thermodynamics\cite{BCH}. This close mathematical analogy was at first conceived to be simply coincidental, however it soon become clear that black holes do indeed behave as thermodynamic systems. The crucial step towards such a perspective was Hawking's remarkable discovery that quantum particle creation effects result in an effective emission of particles from a black hole with a blackbody spectrum at the temperature proportional to the surface gravity of black hole horizon as $T=\frac{\kappa}{2\pi}$\cite{Hawking}. Thus the four laws of black hole mechanics, in essence, are nothing but a description of black hole thermodynamics. In particular, the thermodynamic entropy of a black hole is given by the Bekenstein-Hawking formula as $S=\frac{A}{4}$ with $A$ the area of black hole horizon. Note that the entropy of an ordinary physical system is essentially the logarithm of the number of microscopic states compatible with the observed macroscopic state, therefore the assignment of one quarter of the area as the black hole entropy seems to indicate that holography is the fundamental nature of gravity\cite{Hooft,Susskind}. The breakthrough towards such a holographic gravity was made in 1998 by Maldacena, who demonstrated that the bulk gravity in AdS is dual to the ordinary quantum system on the boundary, where the entropy of boundary system is given by the bulk black hole entropy\cite{Maldacena}. This explicit implementation of holographic gravity is now dubbed as AdS/CFT correspondence\cite{Witten,GKP}.

With this in mind, one is tempted to expect that any universal law for the ordinary physical systems should have its counterpart on the gravity side. In particular, as highlighted by Prigogine, there are two other intriguing universal behaviors associated with the entropy production rate for the ordinary physical systems. The purpose of this essay is to show that these two universal behaviors are also obeyed by the black hole entropy. In the next section, we shall review these two universal behaviors for the ordinary physical systems. Then using Raychaudhuri equation, we shall prove the validity of such universal laws in the black hole physics. We conclude this essay with some discussions in the end.
\section{Prigogine's Argument}\label{Pri}
The second law of thermodynamics states that in any physically allowed process the entropy production rate is always non-negative, i.e.,
\begin{equation}
\dot{S}\geq 0,
\end{equation}
where the dot denotes the time derivative. Furthermore, there is a general extremum principle associated with the entropy production rate, namely the entropy production rate reaches its mimimal value at the steady state, i.e.,
\begin{equation}
\delta\dot{S}=0,\delta^2\dot{S}\geq 0\label{A}
\end{equation}
for the steady state. The general formulation and the demonstration of the validity of this principle is due to Prigogine. In addition, as the system approaches the steady state, the entropy production becomes slower and slower, i.e.,
\begin{equation}
\ddot{S}\leq 0\label{B}
\end{equation}
near the steady state.

Eq.(\ref{A}) and Eq.(\ref{B}) are just the two additional universal behaviors regarding the entropy production rate for the ordinary physical systems\cite{KP}. Note that both of them are concerned with the entropy production rate near the steady state, so one can prove them by the linear response theory. Specifically speaking, in the linear regime the entropy production rate is given by the following formula
\begin{equation}
\dot{S}=\int \mathbf{J}^k\cdot\mathbf{F}_k=\int L^{kj}\mathbf{F}_j\cdot\mathbf{F}_k,
\end{equation}
where $\mathbf{J}$ are the currents and $\mathbf{F}$ are the forces, with the Onsager matrix $L$ semi-positive in accordance with the second law of thermodynamics. Whence we have
\begin{eqnarray}
\delta\dot{S}&=&\int L^{kj}\delta\mathbf{F}_j\cdot\mathbf{F}_k+\int L^{kj}\mathbf{F}_j\cdot\delta\mathbf{F}_k=
2\int\mathbf{J}^k\cdot\delta\mathbf{F}_k=-2\int\mathbf{J}^k\cdot\mathbf{\nabla}\delta f_k\nonumber\\
&=&-2\oint \mathbf{n}\cdot\mathbf{J}^k\delta f_k+2\int\mathbf{\nabla}\cdot\mathbf{J}^k\delta f_k
=-2\oint \mathbf{n}\cdot\mathbf{J}^k\delta f_k-2\int \dot{\rho}^k\delta f_k=-2\oint \mathbf{n}\cdot\mathbf{J}^k\delta f_k\nonumber\\
\end{eqnarray}
where we have used the Onsager reciprocal relation in the second step, the definition for the forces $\mathbf{F}=-\mathbf{\nabla}f$ in the third step, the continuity equation $\dot{\rho}+\mathbf{\nabla}\cdot\mathbf{J}=0$ in the fifth step, and the definition for the steady state in the sixth step. So the variation of entropy production rate vanishes at the steady state, i.e., $\delta\dot{S}=0$ for an isolated system or an open system with $f$ fixed on the boundary. In this case, we further have
\begin{equation}
\delta^2\dot{S}=2\int L^{kj}\delta\mathbf{F}_j\delta\mathbf{F}_k+2\int\mathbf{J}^k\cdot\delta^2\mathbf{F}_k\geq 2\int\mathbf{J}^k\cdot\delta^2\mathbf{F}_k=0
\end{equation}
for the steady state. Regarding the second universal behavior, we have
\begin{eqnarray}
\ddot{S}&=&2\int \mathbf{J}^k\cdot\dot{\mathbf{F}}_k=-2\int \mathbf{J}^k\cdot\mathbf{\nabla}\dot{f}_k=-2\oint \mathbf{n}\cdot\mathbf{J}^k\dot{f}_k-2\int\dot{\rho}^k\dot{f}_k=-2\int\dot{\rho}^k\dot{f}_k,
\end{eqnarray}
 Note that the variation of $\rho$ is related to the variation of $f$ by the Hessian matrix, which is positive due to the thermodynamical stability. Thus we have
\begin{equation}\label{thermo}
\ddot{S}=-2\int H^{kj}\dot{f}_j\dot{f}_k\leq 0.
\end{equation}

\section{Raychaudhuri's Intuition}\label{Ray}
Corresponding to Eq.(\ref{A}) and Eq.(\ref{B}), we are required to check whether
\begin{equation}
\delta\dot{A}=0,\delta^2\dot{A}\geq 0 \label{C}
\end{equation}
 are satisfied at the black hole stationary state, and whether
\begin{equation}
\ddot{A}\leq 0 \label{D}
\end{equation}
as the black hole approaches the final stationary state. To this end, recall that the dynamics of black hole event horizon  is controlled by Raychaudhuri equation as follows\cite{Wald}
\begin{eqnarray}
l^c\nabla_c\hat{\sigma}_{ab}&=&-\theta\hat{\sigma}_{ab}+\widehat{C_{cbad}l^cl^d},\nonumber\\
l^c\nabla_c\theta&=&-\frac{1}{2}\theta^2-\hat{\sigma}^{ab}\hat{\sigma}_{ab}-R_{ab}l^al^b=-\frac{1}{2}\theta^2-\hat{\sigma}^{ab}\hat{\sigma}_{ab},
\end{eqnarray}
where we have used the vacuum Einstein equation with or without a cosmological constant in the second step of the second equation, with $\hat{\sigma}$ the shear of the congruence of null geodesic generators $l^a$ and $\theta$ the expansion. It follows from the second Raychaudhuri equation that the expansion $\theta\geq 0$ everywhere on the black hole horizon because the null generators can never run into caustics\cite{Wald}. By the identity $\theta=\frac{l^a\nabla_a A}{A}$ with $A$ the local area element of black hole horizon, we are thus led to the second law of black hole thermodynamics. In particular, for a stationary black hole, Raychaudhuri equation further implies that that $\widehat{C_{cbad}l^cl^d}$, $\hat{\sigma}_{ab}$, and $\theta$ vanishes everywhere on the horizon.

Now the strategy for our proof is to perturb an initially stationary black hole and see how it eventually settles down to another stationary black hole. To simplify our calculation, we make use of the diffeomorphism freedom in identifying the perturbed spacetime with the unperturbed background spacetime such that the corresponding null geodesic generators coincide with each other\cite{GW}. Thus the perturbed null geodesics can be expressed as $l^a=l_0^a+\epsilon l_1^a$ with $l_1^a\propto l_0^a$. Similarly, we can expand all geometric quantities such as the Weyl tensor as well as the shear and expansion in terms of power series of $\epsilon$ as follows
\begin{eqnarray}
C_{cbad}&=&C_{0cbad}+\epsilon C_{1cbad}+\epsilon^2C_{2cbad}+\cdot\cdot\cdot,\nonumber\\
\hat{\sigma}_{ab}&=&\hat{\sigma}_{0ab}+\epsilon\hat{\sigma}_{1ab}+\epsilon^2\hat{\sigma}_{2ab}+\cdot\cdot\cdot,\nonumber\\
\theta&=&\theta_0+\epsilon\theta_1+\epsilon^2\theta_2+\cdot\cdot\cdot.
\end{eqnarray}
Plugging the above expansion into Raychaudhuri equation, and solving it order by order with the vanishing boundary condition for all the quantities at a very late time\footnote{Here it is implicitly assumed that the black hole in consideration is dynamically stable.}, we have
\begin{equation}
\widehat{C_{0cbad}l_0^cl_0^d}=0,\hat{\sigma}_{0ab}=0,\theta_0=\theta_1=0,
\end{equation}
and
\begin{eqnarray}
\dot{\hat{\sigma}}_{1ab}&=&\widehat{C_{1cbad}l_0^cl_0^d},\nonumber\\
\dot{\theta}_2&=&-\hat{\sigma}_1^{ab}\hat{\sigma}_{1ab}\leq 0,
\end{eqnarray}
where the dot denotes the time derivative with respect to the affine parameter $\lambda$ of the unperturbed null geodesic generators $l_0^a$ and $\hat{\sigma}_1^{ab}$ is obtained by raising $\hat{\sigma}_{1ab}$ with the unperturbed metric. Next plugging the perturbation expansion of the local area element of the black hole horizon
\begin{equation}
A=A_0+\epsilon A_1+\epsilon^2 A_2+\cdot\cdot\cdot,
\end{equation}
into the aforementioned identity $\theta=\frac{l^a\nabla_a A}{A}$, we end up with
\begin{eqnarray}
\dot{A}_0&=&A_0\theta_0=0,\nonumber\\
\dot{A}_1&=&A_0\theta_1+A_1\theta_0=0,\nonumber\\
\dot{A}_2&=&A_0\theta_2+A_1\theta_1+A_2\theta_0=A_0\theta_2\geq 0.
\end{eqnarray}
The last two equations amount to saying that $\delta\dot{A}=0$  and $\delta^2\dot{A}\geq 0$ respectively, because $A_1$ is essentially the first variation of $A$, and $A_2$ is one half of the second variation of $A$. In addition, the third equation further gives rise to
\begin{equation}\label{dyn}
\ddot{A}_2=\dot{A}_0\theta_2+A_0\dot{\theta}_2=A_0\dot{\theta}_2\leq 0,
\end{equation}
which fulfills the proof of Eq.(\ref{D}).

\section{Discussions}
Motivated by the black hole thermodynamics and AdS/CFT correspondence, we have proved that not only do the two universal behaviors apply to the ordinary physical systems but also apply to the black holes. It is noteworthy that our result does not require the black holes to be asymptotically AdS, which together with the four laws of black hole thermodynamics suggests that the holographic nature of gravity is supposed to be universal, not restricted only within the context of AdS/CFT correspondence. In addition, for simplicity, we just work out the explicit strategy for the four dimensional neutral black hole, but nevertheless our proof can be readily extended to more general cases such as charged or hairy black holes in arbitrary dimensions.

Here comes a remark on the differences between the ordinary physical systems and black holes. For one thing, the validity of Eq.(\ref{thermo}) relies on the thermodynamical stability of the ordinary physical systems while the validity of Eq.(\ref{dyn}) relies on the dynamical stability of the black holes in consideration. Unlike the ordinary physical systems, the dynamical stability of the black holes, however, does not generically imply the thermodynamical stability\cite{HW}. But nevertheless our result applies to such black holes as the asymptotically flat Schwarzschild black hole. For another,
as shown for the ordinary physical systems, the two universal behaviors are associated only with the total entropy production rate rather than the local entropy production rate, because the entropy can flow from one place of the system in consideration to another. However, for the black holes, the two universal behaviors are actually valid not only for the total area but also for the local area element. The reason for this may arise in the fact that there is no causal signal transmittable from one null geodesic generator to another.

We conclude this essay with two outlooks. First, it is worthwhile to check whether the two universal laws also hold for the non-Einstein gravity, because in this case, the black hole entropy is not associated with the area of black hole horizon any more, instead given by Wald formula\cite{Wald1,IW}. In addition,
what we have achieved is actually associated only with the equilibrium state of isolated black holes. However, as we know, the two universal laws also hold for the non-equilibrium steady state of the ordinary open systems, so it is interesting to develop the counterpart of such a non-equilibrium steady state on the gravity side and check whether the two universal laws still survive.

\section*{Acknowledgements}
M.G. is partially supported by NSFC with Grant Nos.11235003, 11375026 and NCET-12-0054, as well as by ``the Fundamental Research Funds for the Central Universities" with Grant No.2015NT16.  Y.T. is partially supported by NSFC with Grant No.11475179 and the Opening Project of Shanghai Key Laboratory of High Temperature Superconductors(14DZ2260700). X.W. is supported by NSFC with Grant Nos. 11475179, 11175245 and
11575286.
 H.Z. is supported in part by the Belgian Federal
Science Policy Office through the Interuniversity Attraction Pole
P7/37, by FWO-Vlaanderen through the project
G020714N, and by the Vrije Universiteit Brussel through the
Strategic Research Program ``High-Energy Physics''. He is also an individual FWO Fellow supported by 12G3515N.


\begin{thebibliography}{20}
\bibitem{BCH}J. M. Bardeen, B. Carter, and S. W. Hawking, Commun. Math. Phys. 31, 161(1973).
\bibitem{Hawking}S. W. Hawking, Commun. Math. Phys. 43, 199(1975).
\bibitem{Hooft}G. $^¡¯$t Hooft, arXiv:gr-qc/9310026.
\bibitem{Susskind} L. Susskind, J. Math. Phys. 36, 6377(1995).
\bibitem{Maldacena}J. Maldacena, Adv. Theor. Math. Phys. 2, 231(1998).
\bibitem{Witten} E. Witten, Adv. Theor. Math. Phys. 2, 253(1998).
\bibitem{GKP}S. Gubser, I. R. Klebanov, and A. M. Polyakov, Phys. Lett. B 428, 105(1998).
\bibitem{KP}D. Kondepudi and I. Prigogine, Modern Thermodynamics: From Heat Engines to Dissipative Structures(Wiley, 2015).
\bibitem{Wald}R. M. Wald, General Relativity(University of Chicago Press, 1984).
\bibitem{GW}S. Gao and R. M. Wald, Phys. Rev. D 64, 084020(2001).
\bibitem{HW}S. Hollands and R. M. Wald, Commun. Math. Phys. 321, 629(2013).
\bibitem{Wald1}R. M. Wald, Phys. Rev. D 48, R3427(1993).
\bibitem{IW}V. Iyer and R. M. Wald, Phys. Rev. D 50, 846(1994).

\end{thebibliography}
\end{document}